\documentclass[twocolumn,showpacs,superscriptaddress,9pt]{revtex4-1} %% use 11pt for Applied Optics
\usepackage{amsmath,amssymb,graphicx}
\usepackage[utf8]{inputenc}
\usepackage{flushend,pifont,color}
\begin{document}

\title{Co-design of an in-line holographic microscope with enhanced axial resolution: selective filtering digital holography}
\author{Nicolas Verrier}\email{nicolas.verrier@univ-st-etienne.fr}
\author{Corinne Fournier}
\author{Anthony Cazier}
\author{Thierry Fournel}
\affiliation{Université de Lyon, F-42023 Saint-Etienne, France, CNRS UMR 5516 Laboratoire Hubert Curien, F-42000 Saint-Etienne, France, Université de Saint-Etienne Jean Monnet, F-42000 Saint-Etienne, France.}

			\begin{abstract}
			Common-path digital in-line holography is considered as a valuable 3D diagnostic techniques for a wide range of applications. This configuration is cost effective and relatively immune to variation in the experimental environment. Nevertheless, due to its common-path geometry, the signal to noise-ratio of the acquired hologram is weak as most of the detector (i.e. CCD/CMOS sensor) dynamics is occupied by the reference field signal, whose energy is orders of magnitude higher than the field scattered by the imaged object. As it is intrinsically impossible to modify the ratio of energy of reference to the object field, we propose a co-design approach (Optics/Data Processing) to tackle this issue. The reference to object field ratio is adjusted by adding a 4-f device to a conventional in-line holographic set-up, making it possible to reduce the weight of the reference field while keeping the object field almost constant. Theoretical analysis of the Cràmer-Rao lower bounds of the corresponding imaging model illustrate the advantages of this approach. These lower bounds can be asymptotically reached using a parametric inverse problems reconstruction. This implementation results in a $60\ \%$ gain in axial localization accuracy (for of 100$\ \mu\rm m$ diameter spherical objects) compared to a classical in-line holography set-up.
			
			\end{abstract}
			% REPLACE WITH CORRECT OCIS CODES FOR YOUR ARTICLE
                          % NOTE: \ocis{} IS ALIASED TO \pacs{} BUT MUST
                          % FORMAT THE TERMS CORRECTLY FOR EACH JOURNAL

	\maketitle %% required		
			
			%% Intro
			%%%%%%%%%%%%%%%%%%%%%%%%%%%%%%%%%%%%%%%%%%%
		\section{Introduction}	
			Digital in-line holography is a widely used tool for 3D imaging in severe experimental environments~\cite{Jones1978,Jericho2006,Jericho2010,Berg2011,VerrierFournier2014}. The common-path geometry of a classical in-line holography set-up makes it possible to perform quantitative 3D imaging with a limited number of optical accesses as well as relative immunity to variations in the experimental conditions (e.g. variations in temperature, vibrations in the imaging path). These features have increased interest in the use of digital in-line holography for application in a wide range of domains including fluid dynamics~\cite{Katz2010,Berg2011}, and bio-medical imaging~\cite{Bishara2010,Mudanyali2013}. Nevertheless, as both the reference field and the object field have a common optical path, it is impossible to act on one or the other independently. This issue can be tackled with an off-axis digital holography configuration~\cite{Leith1962}, which, for instance, makes it possible to characterize both amplitude and phase objects~\cite{Cuche1999,Charriere2006} down to shot noise limits~\cite{Lesaffre2013,VerrierAlloulGross2015,Lopes2015}. However, the complexity of the experimental set-up is not suitable for severe experimental environments. 
			
			Several strategies have been proposed to get round this problem. Signal processing based approaches have been demonstrated based on \emph{a priori} knowledge of the object~\cite{Yang2005}, deconvolution strategies~\cite{Latychevskaia2010}, or compressive imaging~\cite{Rivenson2010,Rivenson2013}. Experimental schemes, based on the use of two orthogonal view points, have also been considered at the cost of a more complex experimental set-up and a mandatory view registration~\cite{Buchmann2013}. However, all these approaches rely on hologram back-propagation to retrieve information about the recorded objects, therefore introducing artifacts such as aliasing ghosts or twin-image noise~\cite{Onural2000,Fournier2004,Onural2007}.
			
			Instead of transforming the data through light back-propagation calculations, inverse problems (IP) approaches have been successfully applied to the reconstruction of in-line and off-axis holograms~\cite{SoulezDenisFounier2007,SoulezDenisThiebaut2007,Denis2009,Bourquard2013,Fournier2014}. The aim of these approaches is to find, in the least-square sense, imaging models that best match the acquired data, and allow accurate estimation of the parameters of the objects under investigation (e.g. 3D position and size of a particle in in-line particle holography). It should be noted that a maximum a posteriori (MAP) approach can be used to regularize the maximum likelihood estimation. Despite good accuracy, this device is limited by the signal to noise ratio of the diffraction pattern (i.e. the object signal), which prevents the detection of particles less than 5 $\mu\rm m$ in diameter with a unitary magnification configuration. To improve accuracy, while keeping ease of use of the experimental set-up, one can consider optimizing both the experimental and the data processing design.
			
			Here, we consider an optics/data-processing ``co-design'' scheme to improve the axial accuracy of a conventional in-line holographic set-up, while preserving its experimental advantages (i.e. limited amount of optical accesses, space bandwidth product, and relative robustness to experimental environment variation). The proposed approach is applied, to the particular case of spherical opaque particle holograms. To this end, we propose a modification of in-line holography involving a selective 4-f filtering stage. Associated parametric imaging model is discussed, and the optimal filter design is discussed through a Cràmer-Rao lower bounds analysis. The advantages of the approach are demonstrated through numerical and bench-top experiments resulting in an axial localization accuracy gain of $60\ \%$.

		\section{Principles of selective filtering}
		\subsection{Experimental set-up}
\begin{figure}[h]
		\centering
		\includegraphics[width = 8.4 cm]{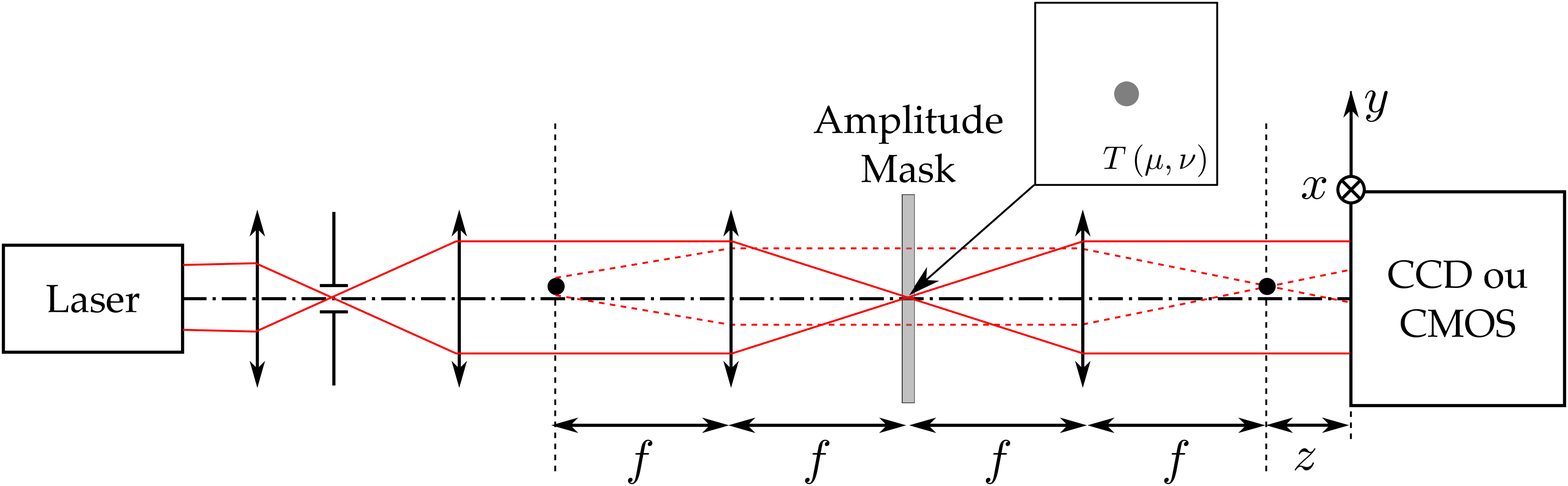}
		\caption{(Color online) Experimental configuration for hologram acquisition.}\label{Fig:SetupStrio}
		\end{figure}

		The experimental set-up for the adjustment of the reference to object field ratio is illustrated Fig. \ref{Fig:SetupStrio}. It consists of a conventional in-line digital holography set-up. The light emitted by a $\lambda=660\ \rm nm$ fiber-pigtailed laser diode (Coherent OBIS 660 LX-FP \textregistered) is collimated to act as the illumination beam.
		Selective filtering of hologram content is performed by a 4-f stage, made of two $f=150\ \rm mm$ achromatic doublets (Thorlabs AC508-150-A-ML \textregistered), which makes it possible to perform all optical Fourier filtering operations. The test object - a chrome deposited opaque disk whose diameter is $2r = 100\ \mu \rm m$ (diameter $\pm 1\ \mu\rm m$, roundness $\pm$ 0.25 $\mu$m Optimask \textregistered) - is positioned in the object focal plane of the 4-f device, and illuminated by a collimated field. Therefore, in the Fourier space of the 4-f arrangement, the reference field is focalized, whereas the wave diffracted by the object spreads in the whole spatial frequency space. Note that this configuration is similar to the one proposed in Ref.~\cite{Mico2009}, where phase shifting in-line holography was considered. This configuration makes it possible to act on the reference field while keeping the object field high-frequency content. The filtering mask, whose optimal design is discussed in the following sub-sections, consists of a semi-transparent disk deposited on a glass substrate. This filtering mask acts on the amplitude of the optical Fourier transform of the interference field: the central part of the Fourier spectrum is attenuated, while the remainder of the spectrum remains unchanged, which allows adjustment of the reference to object field ratio. The filtered Fourier transform is then collected by the second lens, which performs another Fourier transform, leading to a reference attenuated version of our original interference field, which is propagated over a distance $z$ to the recording plane. The interference is finally recorded on a 12 bit $4872\times3248$ pixel with $9\ \mu m$ pitch CCD sensor (Prosilica GE4900 \textregistered).
		
		\subsection{Imaging model}
		To be able to optimally design the filtering mask, we had to build the imaging model associated with our experimental set-up. Considering a conventional in-line holography set-up, the recorded intensity is
	\begin{equation}\label{Eq:Fresnel}
	I_z\left(x,y\right) \propto \left|1-\vartheta\left(x,y\right)*h_z\left(x,y\right)\right|^2,
	\end{equation}
	where $\vartheta$ is the 2D aperture function of the particle under investigation, which is unity within the aperture and zero elsewhere. The symbol $*$ denotes the 2D spatial convolution product, and the impulse response of free-space propagation $h_z\left(x,y\right)$ is
	\begin{equation}\label{Eq:hz}
	h_z\left(x,y\right)=\frac{1}{i\lambda z}\exp\left[i\frac{\pi}{\lambda z}\left(x^2+y^2\right)\right].
	\end{equation}
		In the remainder of this article, $z$ is set so that the assumption $z\gg r^2/\lambda$ holds. Under this assumption, the Huygens-Fresnel diffraction integral finds an analytic expression, often called the Tyler/Thompson model~\cite{Tyler1976}. The intensity $I_z\left(x,y\right)$ recorded for a spherical opaque object located at a distance $z$ from the imaging sensor, and without filtering operations, is thus given by
		\begin{multline}\label{Eq:Thompson}
		I_z\left(x,y\right) \propto 1 - \frac{1}{\lambda z}\mathcal{F}_{\frac{x}{\lambda z},\frac{y}{\lambda z}}\left\{\vartheta\left(\xi,\eta\right)\right\}\sin\left[\frac{\pi}{\lambda z}\left(x^2+y^2\right)\right]\\
		+\left[\frac{1}{\lambda z}\mathcal{F}_{\frac{x}{\lambda z},\frac{y}{\lambda z}}\left\{\vartheta\left(\xi,\eta\right)\right\}\right]^2,
		\end{multline}
		Considering an opaque diffraction particle, the Fourier transform of $\vartheta$ is 
		\begin{multline}\label{Eq:Bessel}
		\mathcal{F}_{\frac{x}{\lambda z},\frac{y}{\lambda z}}\left\{\vartheta\left(\xi,\eta\right)\right\}=2\pi r^2\left(\frac{\lambda z}{2\pi r \sqrt{x^2+y^2}}\right) \\
		\times J_1\left(\frac{2\pi r \sqrt{x^2+y^2}}{\lambda z}\right),
		\end{multline}
		with $J_1$ the Bessel function of the first kind. Introducing the cardinal Bessel function of the first kind $J_{1_c}$, Eq. (\ref{Eq:Bessel}) becomes
		\begin{equation}
		\mathcal{F}_{\frac{x}{\lambda z},\frac{y}{\lambda z}}\left\{\vartheta\left(\xi,\eta\right)\right\}=2\pi r^2 J_{1_c}\left(\frac{2\pi r \sqrt{x^2+y^2}}{\lambda z}\right).
		\end{equation}
		It should be noted that, under the assumption of a diluted particle sample, an additive intensity model can be built from Eq. (\ref{Eq:Thompson}). In this situation, the interference between object waves will not been taken into account. However, a more complete non-linear imaging model can be considered if this situation is encountered.
		
		Coming back to Eq. (\ref{Eq:Fresnel}), it is possible to build an imaging model that accounts for the selective filtering of the field. Considering that both lenses perform an optical Fourier transform of the field distribution in their object focal plane, we can modify Eq. (\ref{Eq:Fresnel}) to
		\begin{multline}\label{Eq:ImagingModelRaw}
		I_{z}\left(x,y\right) \propto \left|\mathcal{F}_{\frac{x}{\lambda f},\frac{y}{\lambda f}}\left[\mathcal{F}_{\frac{x_0}{\lambda f},\frac{y_0}{\lambda f}}\left\{1-\vartheta\left(\xi,\eta\right)\right\}\right.\right.\\
		\left.\left.\times M_{\alpha}\left(\frac{x_0}{\lambda f},\frac{y_0}{\lambda f}\right)\right]*h_z\left(x,y\right)\right|^2, 
		\end{multline}
		with $M_{\alpha}\left(\frac{x_0}{\lambda f},\frac{y_0}{\lambda f}\right)$ the opacity distribution of the filtering mask defined as
		\begin{eqnarray}\label{Eq:Filter}
		M_{\alpha}\left(\frac{x_0}{\lambda f},\frac{y_0}{\lambda f}\right) = \left\{ {\begin{array}{*{20}{ll}}
											{\alpha\ {\rm if}\ \sqrt{\left(\frac{x_0}{\lambda f}\right)^2+\left(\frac{y_0}{\lambda f}\right)^2}\leq\frac{\rho_{\rm M}}{\lambda f}}\\
													{1\ {\rm{otherwise}}}
											\end{array}} \right.,
		\end{eqnarray}
		where $\rho_{\rm M}$ is the physical size of the filtering mask, and $\left(x_0,y_0\right)$ are the spatial coordinates in the Fourier space.
		The expression given in Eq. (\ref{Eq:ImagingModelRaw}) can be further expanded leading to
		\begin{align}\label{Eq:ImagingModelRaw2}
		I_{z}&\left(x,y\right)\propto\nonumber \\
		&\alpha^2-\alpha\frac{2\pi r^2}{\lambda z}J_{1_c}\left(\frac{2\pi r\sqrt{x^2+y^2}}{\lambda z}\right)\sin\left[\frac{\pi}{\lambda z}\left(x^2+y^2\right)\right]\nonumber	\\
		&+\left\{\frac{2\pi r^2}{\lambda z}\alpha\left(1-\alpha\right)J_{1_c}\left(\frac{2\pi r\sqrt{x^2+y^2}}{\lambda z}\right)\sin\left[\frac{\pi}{\lambda z}\left(x^2+y^2\right)\right]\right\}\nonumber\\
		&*J_{1_c}\left(\frac{2\pi \rho_{\rm M}\sqrt{x^2+y^2}}{\lambda f}\right)+ \rm second\ order\ terms,
		\end{align}
		where $\alpha$ is the amplitude transmittance of the filtering mask. Details of the calculations leading to Eq. (\ref{Eq:ImagingModelRaw2}) are provided in Appendix \ref{Sec:ImagingModel}. As $J_{1_c}\left(\frac{2\pi \rho_{\rm M}\sqrt{x^2+y^2}}{\lambda f}\right)$	can be considered to have a compact support (most of its energy is in the low frequency components), Eq. (\ref{Eq:ImagingModelRaw2}) can be further simplified leading to our imaging model
		\begin{align}\label{Eq:ImagingModel}
		I_{z}&\left(x,y\right)\propto\alpha^2-\alpha\frac{2\pi r^2}{\lambda z}J_{1_c}\left(\frac{2\pi r\sqrt{x^2+y^2}}{\lambda z}\right)\sin\left[\frac{\pi}{\lambda z}\left(x^2+y^2\right)\right]\nonumber	\\
		&+\left\{\frac{2\pi r^2}{\lambda z}\alpha\left(1-\alpha\right)J_{1_c}\left(\frac{2\pi r\sqrt{x^2+y^2}}{\lambda z}\right)\sin\left[\frac{\pi}{\lambda z}\left(x^2+y^2\right)\right]\right\}\nonumber\\
		&\times \vartheta_{\rm M}\left(x,y\right) + \rm second\ order\ terms.
		\end{align}
		It can be seen that Eq. (\ref{Eq:ImagingModel}) is a high pass-filtered version of Eq. (\ref{Eq:Thompson}) (further shown in Fig. \ref{Fig:Amplitude}). Considering Eqs. (\ref{Eq:Filter}) and (\ref{Eq:ImagingModel}), it is possible to simulate filtered holograms. In the remainder of this article, second order terms of Eq. (\ref{Eq:ImagingModel}) will be neglected. As a matter of fact, as it is discussed in Appendix \ref{Sec:Hypothesis}, its maximal value (considering our experimental set-up parameter) is at least one order of magnitude lower than the other terms in Eq. (\ref{Eq:ImagingModel}). To be able to acquire holograms in optimal conditions, we will focus on the co-design of the filter.
		
		\subsection{Filter co-design}
		Two parameters need to be accounted for in order to filter our reference field. First, the filter radius $\rho_{\rm M}$ can be chosen so that the imaging model remains simple, while keeping an analytic expression. This aspect will be convenient to analyze the influence of the filtering mask opacity. Then, the filter transmittance is chosen in order to have optimal performance of the imaging system.
		
		\subsubsection{Determination of the filter radius}
		The role of the filtering mask is to act on the Fourier space low frequency components, mainly associated with the reference field, while keeping the high frequency content almost constant. Considering Eq. (\ref{Eq:ImagingModelRaw}), it can be seen that the filtering is performed on the low frequency content of the Fourier transform of the particle transmittance $1-\vartheta\left(x_0,y_0\right)$. Its analytic expression is straightforwardly derived from Eq. (\ref{Eq:Bessel}) and is
		\begin{equation}\label{Eq:Bessel2}
		\mathcal{F}_{\frac{x_0}{\lambda f},\frac{y_0}{\lambda f}}\left\{1-\vartheta\left(x_0,y_0\right)\right\}=\delta_{0,0}\left(\frac{x_0}{\lambda f},\frac{y_0}{\lambda f}\right)-\mathcal{F}_{\frac{x_0}{\lambda f},\frac{y_0}{\lambda f}}\left\{\vartheta\left(x_0,y_0\right)\right\}, 
		\end{equation}
		where $\delta_{0,0}$ is the Dirac distribution. In order to link the filter radius $\rho_{\rm M}$ to observable quantities such as the particle radius, while keeping an analytic image formation model, the filter radius $\rho_{\rm M}$ is chosen so that it extends over the first cancellation of the modulation function given Eqs. (\ref{Eq:Bessel}) and (\ref{Eq:Bessel2}). It should be noted that the choice of the first cancellation is arbitrary: considering that the object field is not disturbed by the filtering mask, every cancellation of the modulation function can be selected to design $\rho_{\rm M}$. Therefore the mask radius $\rho_{\rm M}$ is given by
		\begin{figure}[t]
		\centering
		\includegraphics[width = 8.4 cm]{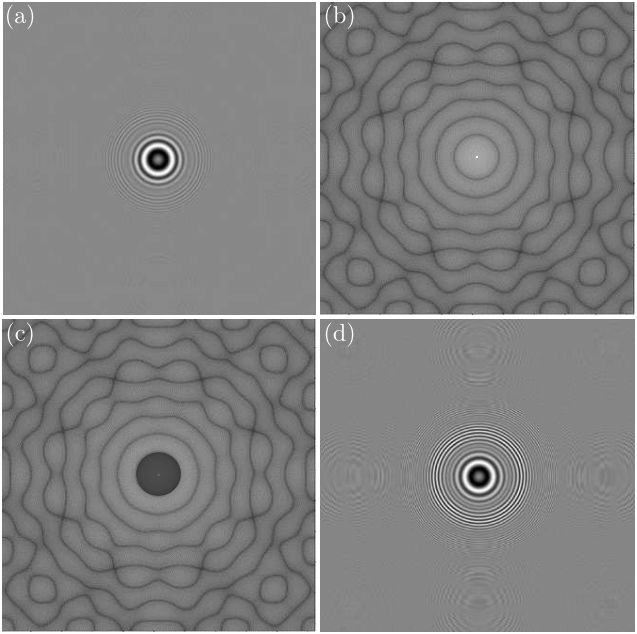}
		\caption{Illustration of the selective filtering. (a) Hologram simulated without filtering. (b) Fourier transform modulus of the particle hologram: $\mathcal{F}_{\frac{x_0}{\lambda f},\frac{y_0}{\lambda f}}\left\{1-\vartheta\left(\xi,\eta\right)\right\}$. (c) Filtered Fourier space: $\mathcal{F}_{\frac{x_0}{\lambda f},\frac{y_0}{\lambda f}}\left\{1-\vartheta\left(\xi,\eta\right)\right\}\times\vartheta_{\rm M}\left(\frac{x_0}{\lambda f},\frac{y_0}{\lambda f}\right)$. Selectively filtered hologram obtained according to Eq. (\ref{Eq:ImagingModel}). (d) Filtered hologram. (b) and (c) are proposed in log-scale for the purpose of illustration.}\label{Fig:Filtrage}
		\end{figure}
		\begin{equation}\label{Eq:Rmasque}
		\rho_{\rm M}=\frac{1.22\lambda f}{2r}.
		\end{equation}
		Thus, according to our experimental configuration, $\lambda = 660\ \rm nm$, $f=150\ \rm mm$, and $r=50\ \mu\rm m$, we obtain $\rho_{\rm M}=1.2\ \rm mm$. The size of the mask is fixed in the remainder of the article. It should be noted that Eq. (\ref{Eq:Rmasque}) can be used to adjust either $f$ or $r$, while keeping $\rho_{\rm M}$ constant. 
		
		Illustration of the proposed scheme is depicted Fig. \ref{Fig:Filtrage}. Here, the effect of the selective filtering is underlined by simulating the hologram, using Eq. (\ref{Eq:ImagingModel}), corresponding to a $r=50\ \mu\rm m$ particle positioned at $z=130\ \rm mm$ from a $1024\times 1024$ sensor, whose pixel pitch is $9\ \mu\rm m$. For the purpose of qualitative comparison, the hologram without filtering, i.e. considering $\alpha=1$, is shown in Fig. \ref{Fig:Filtrage}(a). The field in the filtering plane, which corresponds to the interference light field Fourier transform modulus (Fig. \ref{Fig:Filtrage}(b)) is filtered considering a mask with a radius $\rho_{\rm M}=1.2\ \rm mm$ as predicted by Eq. (\ref{Eq:Rmasque}), and a transmittance $\alpha^{\rm amp}=0.13$, as illustrated in Fig. \ref{Fig:Filtrage}(c). Comparing the unfiltered field (Fig. \ref{Fig:Filtrage}(a)) with its filtered equivalent (Fig. \ref{Fig:Filtrage}(d)) makes it possible to qualitatively assess the advantages of the proposed scheme. As can be seen by comparing the simulations, higher frequencies of the interference signal are more highly contrasted throughout imaging field.
		
		\subsubsection{Influence of the opacity of the filtering mask on the accuracy of parameter estimation} 
		In estimation theory~\cite{Kay1993}, minimal achievable variance of an imaging model parameter can be assessed through Cramer-Rao Lower Bounds (CRLB) computation~\cite{Cramer1999,Rao1992,Refregier2004,Fournier2010}. These bounds can be estimated by inverting the Fisher information matrix associated with the imaging model concerned. Under the assumption of Gaussian white noise, and considering an imaging model relying on four parameters $\theta_{i=1,4}$, the Fisher information matrix is a $4\times 4$ matrix whose elements are proportional to the imaging model gradients along the parameter direction. Considering $f$, $\rho_{\rm M}$ and $\alpha$ as fixed parameters, the Fisher information matrix associated to the imaging model of Eq. (\ref{Eq:ImagingModel}) is given by~\cite{Fournier2010}
		\begin{equation}\label{Eq:Fisher}
		{F_{{\rm{M}}_{i,j}}} = {\rm{SN}}{{\rm{R}}^2} \times \sum_x {\sum_y {{{{\frac{{\partial I^n\left(x,y\right)}}{{\partial {\theta _i}}}\frac{{\partial I^n\left(x,y\right)}}{{\partial {\theta _j}}}} }}} },
		\end{equation}
		\begin{figure}[b]
		\centering
		\includegraphics[width = 8.9 cm]{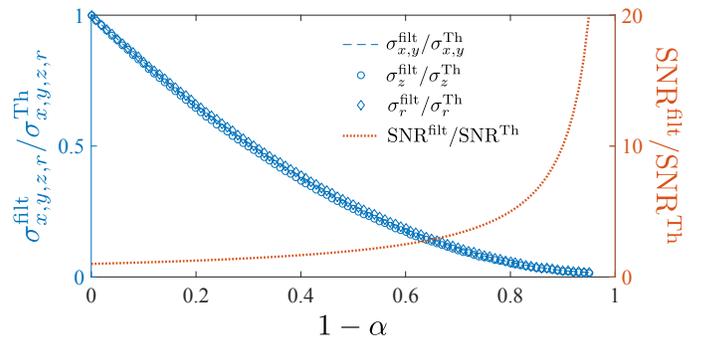}
		\caption{(Color online) Evolution of the hologram SNR (orange dotted line), and the standard deviation of $x,y$ (dashed blue line), $z$ (blue circles), and $r$ (blue diamonds) as a function of the filtering mask opacity $1-\alpha$. All the parameters are normalized to the values of the classical imaging model (see Eq. (\ref{Eq:Thompson})).}\label{Fig:CRLBNum}
		\end{figure}
		where SNR stands for the hologram signal to noise ratio, and $\theta\in\left\{x_i,y_i,z_i,r_i\right\}$ is the set of parameters of the imaging model. The centered and normalized imaging model $I^n$ is given by either Eq. (\ref{Eq:Thompson}) or Eq. (\ref{Eq:ImagingModel}). In the case of uncorrelated imaging model parameters, the non-diagonal terms vanish. The CRLBs on the imaging model can finally be obtained by inverting the Fisher information matrix (Eq. (\ref{Eq:Fisher})). These bound, denoted $\sigma_{x,y,z,r}$, correspond to the diagonal elements of the inverted matrix. Under the assumption of non-correlated imaging model parameter, the CRLB are given by
		\begin{equation}\label{Eq:CRLB}
		F_{\rm M}^{-1}=\left(\begin{matrix}
		\sigma_x^2 & \varepsilon_{yx} & \varepsilon_{zx} & \varepsilon_{rx} \\ 
	\varepsilon_{xy} & \sigma_y^2 & \varepsilon_{zy} & \varepsilon_{ry} \\ 
		\varepsilon_{xz} & \varepsilon_{yz} & \sigma_z^2 & \varepsilon_{rz} \\ 
		\varepsilon_{xr} & \varepsilon_{yr} & \varepsilon_{zr} & \sigma_r^2
		\end{matrix} \right),
		\end{equation}
		with $\varepsilon_{ij}\ll \sigma_{x,y,z,r}$. Under this assumption, the diagonal terms of $F_{\rm M}^{-1}$ can be written as
		\begin{equation}\label{Eq:CRLB2}
		\sigma_i^2\approx \frac{1}{F_{{\rm{M}}_{i,i}}}=\frac{1}{\rm SNR^2}\left[\sum_x\sum_y\left|\frac{\partial I^n\left(x,y\right)}{\partial \theta_i}\right|^2\right]^{-1}.
		\end{equation}
		To compare the accuracy of the conventional digital holography set-up (relying on the imaging model given in Eq. (\ref{Eq:Thompson})) with that of our proposed experimental arrangement (associated with the imaging model of Eq. (\ref{Eq:ImagingModel})), we study the ratio between the standard deviation on the parameter estimation for both configurations
		\begin{equation}\label{Eq:Ratio}
		\frac{\sigma_i^{\rm filt}}{\sigma_i^{\rm Th}}=\frac{\rm SNR^{\rm Th}}{\rm SNR^{\rm filt}}\left(\frac{\sum_x\sum_y\left|\frac{\partial I^{\rm Th}}{\partial \theta_i}\right|^2}{\sum_x\sum_y\left|\frac{\partial I^{\rm filt}}{\partial \theta_i}\right|^2}\right)^{1/2},
		\end{equation}				
		where the superscipts $\rm Th$, and $\rm filt$ are respectively associated with the image formation models given in Eqs. (\ref{Eq:Thompson}) and (\ref{Eq:ImagingModel}). It should be noted that the image of the model gradients differs only in the center. In the case of an attenuation of the first diffraction lobe, the change in the gradient energy impacts less than 4 $\%$ of the pixels. Furthermore, the change occurs in the low frequency region where energy of the gradient is low. Thus, the ratio of the gradient energies is close to unity, and Eq. (\ref{Eq:Ratio}) becomes
		\begin{equation}\label{Eq:AccuracyEvolution}
		\frac{\sigma_i^{\rm filt}}{\sigma_i^{\rm Th}}\approx\frac{\rm SNR^{\rm Th}}{\rm SNR^{\rm filt}}
		\end{equation}
		Therefore, to characterize the effect of filtering parameters estimation on the accuracy gain, Eq. (\ref{Eq:AccuracyEvolution}) is computed for $\alpha\in\left[0,1\right]$. The filtering procedure will mainly affect the hologram SNR. Therefore, considering Eq. (\ref{Eq:CRLB2}), the estimation accuracy will be improved accordingly.
		Fig. \ref{Fig:CRLBNum} shows the results obtained with the evolution of $\sigma_{x,y}^{\rm filt}/\sigma_{{x,y}}^{\rm Th}$ (dashed blue line), $\sigma_z^{\rm filt}/\sigma_{{z}}^{\rm Th}$ (blue circles), and $\sigma_r^{\rm filt}/\sigma_{{r}}^{\rm Th}$ (blue diamonds). The hologram SNR was estimated considering calculations derived in Sec. \ref{Sec:SecD}.  As expected, the greater the opacity of the filtering mask, the higher the hologram SNR (see dashed orange curve in Fig. \ref{Fig:CRLBNum}), and hence the better the parameter estimation.
		It should be noted that this theoretical analysis does not account for experimental parameters such as maximal light source energy, imaging sensor dynamics and quantization, which are likely to prevent from the use of a filtering mask with $\alpha\rightarrow 0$. Nevertheless, it emphasizes the fact that filtering most of the reference beam contribution improves the accuracy of imaging model parameters.

		\subsection{Improving the signal to noise ratio}\label{Sec:SecD}
		As stated earlier, the proposed filtering improves the contrast of high frequency features of the recorded hologram. This effect can be proved experimentally by recording holograms with and without filtering. Considering the imaging model without filtering (see Eq. (\ref{Eq:Thompson}), the maximal intensity $I_z^{\rm max}$ in the recording plane can be written as
		\begin{figure}[t]
		\centering
		\includegraphics[width = 7.5 cm]{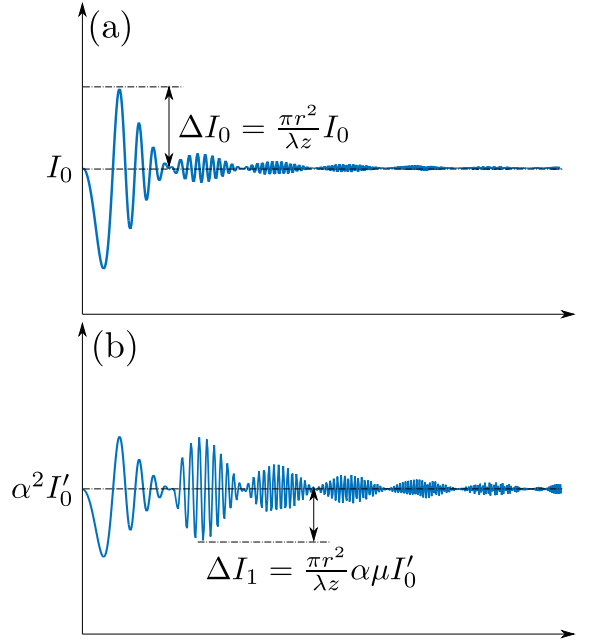}
		\caption{(Color online) Schematics of the recorded intensity profiles (a) without, and (b) with filtering. These schematics are not to scale, as the aim is simply to better highlight the notations introduced in the main matter.}\label{Fig:Amplitude}
		\end{figure}
		\begin{equation}\label{Eq:OffsetThompson}
		I_z^{\rm max}=I_0+\Delta I_0,
		\end{equation}
where $I_0$ is the reference wave offset and $\Delta I_0$ is the amplitude of the first diffraction lobe given by
		\begin{equation}\label{Eq:DI0}
		\Delta I_0=\frac{\pi r^2}{\lambda z} I_0.
		\end{equation}
		The introduced notations are highlighted in Fig. \ref{Fig:Amplitude}. In the same way, the maximal recorded intensity with filtering $I_{\rm filt}^{\rm max}$ can be described as the sum of an offset $I_1$ and an amplitude $\Delta I_1$ so that
		\begin{equation}\label{Eq:OffsetStrio}
		I_{\rm z_{filt}}^{\rm max}=I_1+\Delta I_1,
		\end{equation}
		with $I_1=\alpha^2 I_0'$, and $\Delta I_1$ given by 	
		\begin{equation}
		\Delta I_1=\frac{\pi r^2}{\lambda z}\alpha\mu I_0'.
		\end{equation}
		\begin{figure}[h]
		\centering
		\includegraphics[width = 8.4 cm]{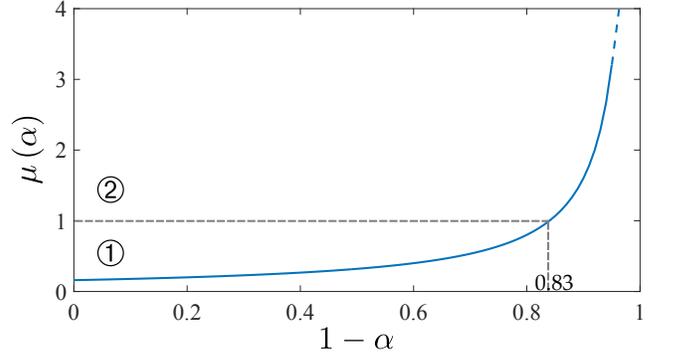}
		\caption{(Color online) Evolution of the ratio $\mu$ between the second and the first diffraction lobe as a function of the opacity of the filtering mask $1-\alpha$.}\label{Fig:MuAlpha}
		\end{figure}
		The ratio $\mu$ between the second and the first maxima of a cardinal Bessel function of the first kind has been added to account for the reduction of the offset, as well as the evolution of the hologram SNR according to $\alpha$.
		In practice, to compensate for the decrease in the signal attributed to the filtering process, the intensity of the laser is adjusted to cover the whole camera dynamic range
		\begin{equation}\label{Eq:AjustementPeche}
		I_{\rm z_{filt}}^{\rm max}=I_z^{\rm max}.
		\end{equation}
		According to the value of the ratio $\mu$, two scenarios, scenario \ding{192} and scenario \ding{193} in Fig. \ref{Fig:MuAlpha}, can be considered:
			\begin{enumerate}
			\item if $\mu\leq 1$ the first diffraction lobe, corresponding to the first maximum of the cardinal Bessel function of the first kind, remains preponderant, the filtering process will only slightly improve the high frequency content of the recorded hologram.
			\item if $\mu>1$ the second diffraction lobe becomes predominant, there will be a marked improvement in the high frequency features in the recorded hologram.
			\end{enumerate}		 
As illustrated in Fig. \ref{Fig:MuAlpha}, the switch between the scenarios takes place when the opacity of the mask is such that $1-\alpha=0.83$. This value was estimated by calculating the ratio $\mu$ between the two first diffraction lobes according to Eq. (\ref{Eq:ImagingModel}). Considering these aspects, and using Eqs. (\ref{Eq:OffsetThompson}) to (\ref{Eq:AjustementPeche}) it is possible to link both $I_0$ and $I_0'$ to fulfill Eq. (\ref{Eq:AjustementPeche})	leading to
		\begin{equation}\label{Eq:RatioIntensity}
		I_0'=\frac{1+\frac{\pi r^2}{\lambda z}}{\alpha\left(\alpha + \frac{\mu\pi r^2}{\lambda z}\right)}I_0.
		\end{equation}
The same can be done to obtain the amplitude ratio between the filtered and unfiltered hologram
		\begin{equation}\label{RatioAmpl}
		\frac{\Delta I_1}{\Delta I_0}=\mu\frac{1+\frac{\pi r^2}{\lambda z}}{\alpha+\mu\frac{\pi r^2}{\lambda z}}
		\end{equation}
		\begin{figure}[t]
		\centering
		\includegraphics[width = 8.4 cm]{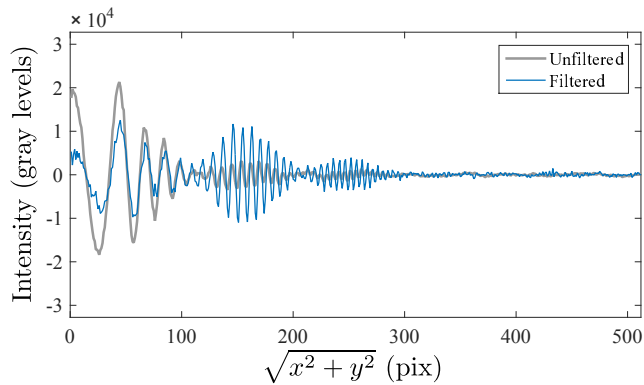}
		\caption{(Color online) Radial intensity profile of the recorded hologram without (gray solid curve), and with (blue solid curve) fitering for $1-\alpha=0.85$ (data average has been removed from the intensity profile curves).}\label{Fig:FilteringEffect}
		\end{figure}
Fig. \ref{Fig:FilteringEffect} gives an example of an intensity profile recorded with $\alpha$ so that the intensity of the second diffraction is close to that of the first diffraction lobe. The unfiltered intensity profile corresponds to the light gray curve. The second to first amplitude ratio is $A_{2}^{\rm nd}/A_{1}^{\rm st}\approx 0.13$ and the SNR of the recorded pattern is driven by both the offset $I_0$ and the maximum amplitude of the signal $I_z^{\rm max}$. In this configuration, the SNR is fixed. In the filtering configuration (blue curve in Fig. \ref{Fig:FilteringEffect}), the ratio $A_{2}^{\rm nd}/A_{1}^{\rm st}$ can be adjusted according to $\alpha$. When $\alpha$ is chosen such that $A_{2}^{\rm nd}/A_{1}^{\rm st}\geq 1$, the SNR of the recorded hologram is driven by the amplitude of the second diffraction lobe, thus enhancing the high frequency content of the hologram, as can be seen in Figs. \ref{Fig:Filtrage} and \ref{Fig:CompStrio}. In this configuration $1-\alpha=0.85$, and the ratio $\mu=A_{2}^{\rm nd}/A_{1}^{\rm st}\approx 1$, which is in good agreement with the imaging model analysis proposed in Fig. \ref{Fig:MuAlpha}. Moreover, the SNR of the filtered hologram (estimated has the ratio between the second diffraction lobe signal and the noise variance) has been estimated to be 5.19 times higher than that of the unfiltered hologram. This is in accordance with the results obtained in Fig. \ref{Fig:CRLBNum}. Adjusting the hologram SNR makes it possible to act on the estimation accuracy of our imaging model parameters as discussed through our CRLB analysis: the higher the SNR, the better the parameter estimation accuracy. It should however be noted that a sufficient part of the reference beam has to be kept in order to record interference and therefore to function in a holographic regime.
		
		\section{Experimental results}
		The qualitative improvement of the filtering scheme presented above was confirmed by the simulation results. For experimental validation, a custom made filtering mask (transmittance in intensity $\alpha=0.1\pm\ 15\%$ Optimask \textregistered) with radius $\rho_{\rm M}=1.2\ \rm mm$ was positioned within the 4-f filtering stage. The corresponding mask amplitude transmittance was $\alpha=0.32$.
A hologram of a $2r=100\ \mu\rm m$ opaque circular disk was recorded with and without the filtering mask in the 4-f stage. The image of the object through the 4-f arrangement was positioned at $z=130\ \rm mm$ from the sensor whose area was cropped to $1024\times 1024$ pixels for the purpose of illustration.

		\begin{figure}[h]
		\centering
		\includegraphics[width = 6 cm]{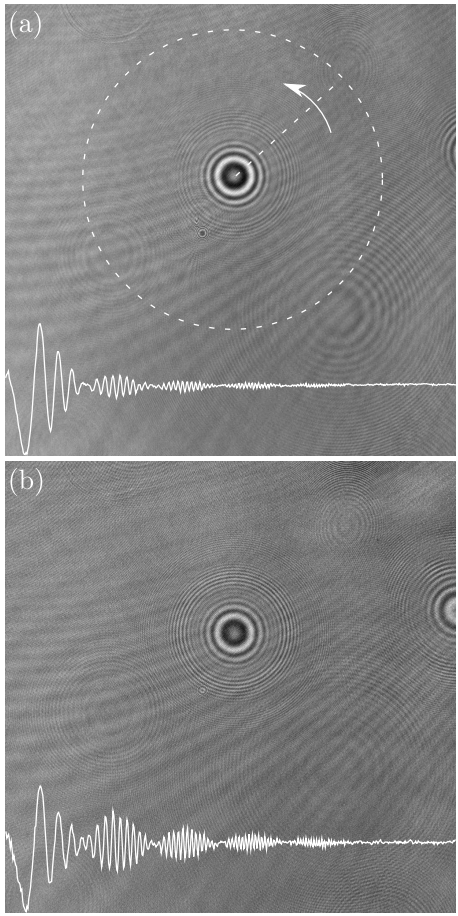}
		\caption{(Color online) Experimental illustration of the filtering effect. (a) Hologram recorded without the filtering mask. (b) Hologram recorded with the filtering mask. Intensity profiles, in white, were obtained by radially averaging the intensity profiles along the circumference of the interference pattern.}\label{Fig:CompStrio}
		\end{figure}
Example of holograms are shown in Fig. \ref{Fig:CompStrio} without (a), and with (b) filtering. The addition of the filtering mask improves the high frequency content contrast. This aspect is made clearer by calculating radial means around the interference pattern. These are shown in red, in Figs. \ref{Fig:CompStrio}(a) and (b).

To better point out the improvement of the proposed scheme on the accuracy of $z$ and $r$ estimation, a statistical experiment was performed. Statistical series of holograms were acquired with the experimental arrangement illustrated in Fig. \ref{Fig:SetupStrio}. Series of 100 holograms, randomly shifted using a mechanical $\left(x,y\right)$ translation stage, were acquired without and with the filtering mask ($\alpha=0.32$ in amplitude) in the 4-f filtering stage. Holograms were then reconstructed using an Inverse Problems (IP) reconstruction scheme~\cite{SoulezDenisFounier2007,SoulezDenisThiebaut2007}. This approach aims at finding, in the least-square sense, the imaging model that best matches the recorded data. For parametric imaging models relying on a few parameters (i.e. object 3D positions and size), these approaches lead to almost unsupervised algorithms, and optimal signal reconstruction~\cite{Mortensen2010,Fournier2010}.

The IP algorithm used in this study is derived from that of~\cite{SoulezDenisFounier2007,SoulezDenisThiebaut2007} and consists of two steps:
\begin{enumerate}
\item A coarse estimation step, aimed at finding the best-matching element in a discrete model dictionary, built by varying the imaging model parameters $\left\{x_i,y_i,z_i,r_i\right\}$. Discrete model dictionaries were built for non-filtered and filtered holograms using Eqs. (\ref{Eq:Thompson}) and (\ref{Eq:ImagingModel}) respectively.
\item A local optimization step, which, with a sub-pixel accuracy, fits the imaging model using, as first guess, the coarsely estimated parameters (see step 1).
\end{enumerate}
		\begin{figure}[t]
		\centering
		\includegraphics[width = 8.4 cm]{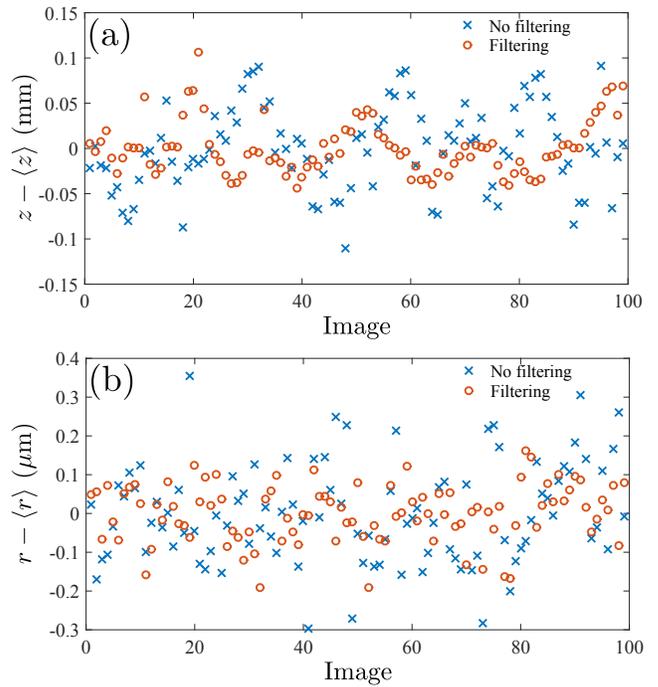}
		\caption{(Color online) Evolution of the estimated object to sensor distance (a) and object radius (b) as a function of the fram index. Blue crosses correspond to the non-filtered filtered holograms that are processed using Eq. (\ref{Eq:Thompson}) as an imaging model, and red circles correspond to  the filtered holograms (associated imaging model corresponding to Eq. (\ref{Eq:ImagingModel})).}\label{Fig:Dispersion_z_r}
		\end{figure}

Unfiltered holograms were processed using the imaging model proposed in Eq. (\ref{Eq:Thompson}), whereas the filtered holograms relied on the imaging model proposed in Eq. (\ref{Eq:ImagingModel}). For each hologram, we thus obtained the object's lateral positions $\left\{x,y\right\}$, axial location $z$, as well as its radius $r$. Doing this for all the filtered and unfiltered holograms makes it possible to extract the experimental statistics concerning parameter estimation accuracy. Due to the random nature of the imposed lateral shift, improvement in lateral accuracy is not discussed here rather we focus our attention on the improvement in accuracy of the $z$ and $r$ parameters, which are assumed to remain constant over the whole hologram sequence. Note that joint optimization of the hologram sequence, discussed in~\cite{VerrierFournier2015,VerrierFournierFournel2015} but not implemented here, can be used to further improve parameter estimation accuracy.

The results obtained are summarized in Fig. \ref{Fig:Dispersion_z_r} for the evolution of $z$ (Fig. \ref{Fig:Dispersion_z_r}(a)), and $r$ (Fig. \ref{Fig:Dispersion_z_r}(b)).  Reconstruction of the filtered hologram is represented by red circles (the corresponding imaging model is given in Eq. (\ref{Eq:ImagingModel})), and blue crosses represent unfiltered holograms (see. Eq. (\ref{Eq:Thompson}) for the reconstruction imaging model). It can be seen that both $z$ and $r$ reconstructed values are less dispersed when selective filtering is performed (see data circled in red in Fig. \ref{Fig:Dispersion_z_r}). This can be confirmed by estimating the standard deviation $\sigma_{z,r}$ and the standard error of the mean $\sigma_{\overline{z,r}}$ of $z$ and $r$ data-sets.
	\begin{table}[htbp]
	\centering
	\caption{\bf Standard deviation and standard error of the mean (estimated over 100 samples) of the axial position $z$ and radius $r$ estimation for both classical and filtered in-line holography}
	\begin{tabular}{ccccc}
	\hline
	 & $\sigma_z\ \left(\mu\rm m\right)$ & $\sigma_{\overline{z}} \left(\mu\rm m\right)$ & $\sigma_r\ \left(\rm nm\right)$ & $\sigma_{\overline{r}}\ \left(\rm nm\right)$ \\
	\hline
	In-line holgraphy & 46.7 & 0.47 & 127.8 & 1.29\\
	Filtering & 29.6 & 0.29 & 75.2 & 0.75\\
	\hline
	Ratio & 1.58 & 1.58 & 1.7 & 1.7 \\
	\hline
	\end{tabular}
	  \label{Tab:Results}
	\end{table}
Results, for in-line holography (corresponding to the set-up proposed Fig. \ref{Fig:SetupStrio} without a filtering mask) and filtered in-line holography are summarized in Table \ref{Tab:Results}. The filtering mask parameters were chosen based on the CRLB analysis. Within this configuration, the average SNR of the hologram sequence was estimated to be 3.3 times higher for the filtered holograms, revealing a good accordance with the theoretical results depicted in Fig. \ref{Fig:CRLBNum}. As can be seen from the standard deviation and from the standard error of the mean on the estimation of both $z$ and $r$, the proposed experimental modification allows an improvement in the accuracy of both parameters, namely $\sigma_{z}^{\rm Th}/\sigma_{z}^{\rm filt}=1.58$ and $\sigma_{r}^{\rm Th}/\sigma_{r}^{\rm filt}=1.7$, thus illustrating the ability of the proposed filtering scheme to enhance the axial localization accuracy of a conventional in-line holography set-up. It should be noted that this enhancement is well bellow the theoretical prediction (rarely reached) in Fig. \ref{Fig:CRLBNum}. This can be explained by the Gaussian white noise hypothesis, not experimentally fulfilled, made in our IP reconstruction scheme that may lead to overestimation of the CRLB values of the imaging model parameters. 

\section{Conclusion}
An optics/data-processing ``co-design'' scheme is proposed as a way to simultaneously improve the object sizing, and axial localization accuracy of a conventional in-line holographic configuration. This approach, based on a close interaction between the experimental set-up and the data reconstruction scheme, is based on a slight modification of a conventional Gabor holographic set-up allowing selective filtering of the reference field while keeping the object field almost intact. This approach has several advantages in terms of the sensor's dynamic range allocation, detection signal to noise ratio, and pattern matching accuracy.

For optimal functioning, the design of the filtering mask has been discussed, through CRLB analysis of the imaging model, leading to a clever choice of the mask opacity. The concept of selective filtering in-line digital holography was qualitatively validated by both digital and bench-top experiments confirming the interest of the method for SNR, and high frequency content improvement of recorded holograms. Quantitative assessment was made through the statistical analysis of 100 holograms with and without selective filtering showing a $60\ \%$ improvement in the accuracy of the axial localization.

Other experimental studies are now needed to identify the limits of the improvement in accuracy with higher mask opacity. Complementary experiments should also be performed on more realistic 3D samples (e.g. droplet jets, colloidal suspension \ldots). Thanks to improvement in SNR, our configuration should make it possible to reconstruct smaller objects.

It should be noted that such an optics/data-processing ``co-design'' methodology is not limited to the particular case of digital holography but can be generalized to any imaging technique that relies on a known parametric imaging model. For more complex objects (non spherical, transparent, phase objects) the herein discussed method can generalized considering a MAP reconstruction~\cite{Denis2009}. This approach thus paves the way for the development of cost effective imaging devices with optimal performance.

The authors acknowledge financial support provided by \emph{Universit\'e de Lyon} through its Programs \emph{``Investissements d'Avenir"} (ANR-1 1-IDEX-0007), and LABEX PRIMES (ANR-11-LABX-0063).

\appendix

\section{Calculation of the imaging model}\label{Sec:ImagingModel}

Details of the calculations leading to the imaging model given in Eq. (\ref{Eq:ImagingModel}) are given here. The amplitude of the light-field in the object plane $U\left(x,y,0\right)$ (see Fig. \ref{Fig:SetupStrio} for details) can be expressed as
		\begin{equation}\label{Eq:ObjectAmplitude}
		U\left(x,y,0\right)=A_0\left[1-\vartheta\left(x,y\right)\right],
		\end{equation}
		where $A_0$ is the constant reference beam amplitude. In the remainder of this Appendix, $A_0$ will be $A_0=1$. The object is positioned in the object focal plane of the 4-f filtering device. The first lens therefore performs an optical Fourier of the amplitude distribution given in Eq. (\ref{Eq:ObjectAmplitude})
		\begin{equation}\label{Eq:TFObjectAmplitude}
		\hat{U}\left(k_x,k_y\right)=\frac{\exp\left(i\frac{2\pi}{\lambda}f\right)}{i\lambda f}\mathcal{F}_{k_x,k_y}\left[U\left(x,y,0\right)\right],
		\end{equation}
		where $\left(k_x,k_y\right)=\left(x/\left(\lambda f\right),y/\left(\lambda f\right)\right)$ are the coordinates in the filtering plane. The Fourier transform of the aperture function $\mathcal{F}_{k_x,k_y}\left[\vartheta\left(x,y\right)\right]$ is given by~\cite{Goodman}
		\begin{equation}\label{Eq:TFAperture}
		\mathcal{F}_{k_x,k_y}\left[\vartheta\left(x,y\right)\right]=\frac{2\pi r^2}{i\lambda f}J_{1_c}\left(2\pi r\sqrt{k_x^2+k_y^2}\right).
		\end{equation}
		As we are only interested in the recorded intensity, the constant phase terms of Eq. (\ref{Eq:TFAperture}) have been omitted.
Introducing Eq. (\ref{Eq:TFAperture}) in Eq. (\ref{Eq:TFObjectAmplitude}) leads to
		\begin{equation}\label{Eq:TFObjectAmplitudeExpanded}
		\hat{U}\left(k_x,k_y\right)=\frac{1}{i\lambda f}\delta_{0,0}\left(k_x,k_y\right)-\frac{2\pi r^2}{i\lambda f}J_{1_c}\left(2\pi r\sqrt{k_x^2+k_y^2}\right),
		\end{equation}
		$\delta_{0,0}\left(k_x,k_y\right)$ being a Dirac distribution. A filtering mask $M_{\alpha}\left(k_x,k_y\right)$ is applied to the amplitude distribution given Eq. (\ref{Eq:TFObjectAmplitudeExpanded}). This amplitude mask is designed so that
		\begin{eqnarray}\label{Eq:AmplitudeMask}
M_\alpha\left(k_x,k_y\right) = \left\{ {\begin{array}{*{20}{ll}}
											{\alpha\ {\rm if}\ \sqrt{k_x^2+k_y^2}\leq \frac{\rho_{\rm M}}{\lambda f}}\\
													{1\ {\rm{otherwise}}}
											\end{array}} \right.,
		\end{eqnarray}
		and the amplitude mask radius $\rho_{\rm M}$ is calculated according to Eq. (\ref{Eq:Rmasque}). The filtered amplitude distribution $\hat{U}_{\rm F}\left(k_x,k_y\right)$ thus becomes
		\begin{eqnarray}
		\hat{U}_{\rm F}\left(k_x,k_y\right)=\frac{\alpha}{i\lambda f}\delta_{0,0}\left(k_x,k_y\right)-\frac{2\pi r^2}{i\lambda f}J_{1_c}\left(2\pi r\sqrt{k_x^2+k_y^2}\right)\nonumber \\
		\times M_{\alpha}\left(k_x,k_y\right),
		\end{eqnarray}

		The second lens of the 4-f filtering device again performs a Fourier transform of the field given in Eq. (\ref{Eq:TFObjectAmplitudeExpanded}). The field in the image focal plane of the second lens is therefore given by (omitting the constant phase terms)
		\begin{equation}
		U\left(x,y,4f\right)=i\lambda f\mathcal{F}_{-k_x\lambda f,-k_y\lambda f}\left[\hat{U}_{\rm F}\left(k_x,k_y\right)\right],
		\end{equation}
		which simplifies as
		\begin{multline}\label{Eq:AmplitudeFielteredField}
		U\left(x,y,4f\right)=\alpha-2\pi r^2\mathcal{F}_{-k_x\lambda f,-k_y\lambda f}\left[J_{1_c}\left(2\pi r\sqrt{k_x^2+k_y^2}\right)\right.\\
		\left.\times M_{\alpha}\left(k_x,k_y\right)\right].
		\end{multline}
		The minus sign in the spatial coordinates $\left(x=-k_x \lambda f,y=-k_y\lambda f\right)$ makes it possible to account for the Gouy phase shift that occurs in the 4-f filtering device~\cite{Gouy}.
		
		Let us now introduce the geometrical aperture of the filtering mask $\vartheta_{\rm M}\left(k_x,k_y\right)$, which can be defined as
		\begin{eqnarray}\label{Eq:FilteringMask}
\vartheta_{\rm M}\left(k_x,k_y\right) = \left\{ {\begin{array}{*{20}{ll}}
											{1\ {\rm if}\ \sqrt{k_x^2+k_y^2}\leq\frac{\rho_{\rm M}}{\lambda f}}\\
													{0\ {\rm{otherwise}}}
											\end{array}} \right.,
		\end{eqnarray}
		allowing Eq. (\ref{Eq:AmplitudeMask}) to be modified as
		\begin{equation}
		M_{\alpha}\left(k_x,k_y\right)=1-\left(1-\alpha\right)\vartheta_{\rm M}\left(k_x,k_y\right).
		\end{equation}
		Introducing $M_{\alpha}\left(k_x,k_y\right)$ in Eq. (\ref{Eq:AmplitudeFielteredField}) makes it possible to derive the amplitude in the 4-f device image plane
		\begin{multline}\label{Eq:FinalAmplitudeObject}
		U\left(x,y,4f\right)=\alpha-\vartheta\left(x,y\right)+\left(1-\alpha\right)\\
		\times \left[\vartheta\left(x,y\right)\underset{k_x,k_y}{*}J_{1_c}\left(2\pi\rho_{\rm M}\sqrt{k_x^2+k_y^2}\right)\right].
		\end{multline}

		Propagation of Eq. (\ref{Eq:FinalAmplitudeObject}) to the sensor plane can be achieved through spatial convolution according to
		\begin{equation}
		U\left(x,y,z\right)=U\left(x,y,4f\right)\underset{x,y}{*}h_z\left(x,y\right),
		\end{equation}
with $h_z\left(x,y\right)$ the free space impulse response defined in Eq. (\ref{Eq:hz}). The amplitude in the sensor plane can be written as	
		\begin{multline}
		U\left(x,y,z\right)=\alpha-\vartheta\left(x,y\right)\underset{x,y}{*}h_z\left(x,y\right)\\
		+\left(1-\alpha \right)\left[\vartheta\left(x,y\right)\underset{x,y}{*} h_z\left(x,y\right)\right]
		\underset{k_x,k_y}{*}J_{1_c}\left(2\pi\rho_{\rm M}\sqrt{k_x^2+k_y^2}\right).
		\end{multline}

Finally, the intensity in the sensor plane can be estimated considering
		\begin{align}\label{Eq:SensorIntensity}
		I_z\left(x,y\right)&=\left|U\left(x,y,z\right)\right|^2\nonumber\\
		&=\left|\mathcal{R}\left(x,y\right)\right|^2-2\Re\left\{\mathcal{R}^*\left(x,y\right)\mathcal{O}\left(x,y\right)\right\}\nonumber\\
		&+\left|\mathcal{O}\left(x,y\right)\right|^2,
		\end{align}
where $\mathcal{R}$ and $\mathcal{O}$ respectively denote the reference and object field defined by
		\begin{equation}
		\mathcal{R}\left(x,y\right)=\alpha,
		\end{equation}
and
		\begin{multline}
		\mathcal{O}\left(x,y\right)=-\vartheta\left(x,y\right)\underset{x,y}{*}h_z\left(x,y\right)\\
		+\left(1-\alpha \right)\left[\vartheta\left(x,y\right)\underset{x,y}{*} h_z\left(x,y\right)\right]
		\underset{k_x,k_y}{*}J_{1_c}\left(2\pi\rho_{\rm M}\sqrt{k_x^2+k_y^2}\right).
		\end{multline}			
		Thus, Eq. (\ref{Eq:SensorIntensity}) can be written as
		\begin{align}
		I_{z}&\left(x,y\right)\propto\alpha^2-\alpha\frac{2\pi r^2}{\lambda z}J_{1_c}\left(\frac{2\pi r\sqrt{x^2+y^2}}{\lambda z}\right)\sin\left[\frac{\pi}{\lambda z}\left(x^2+y^2\right)\right]\nonumber	\\
		&+\left\{\frac{2\pi r^2}{\lambda z}\alpha\left(1-\alpha\right)J_{1_c}\left(\frac{2\pi r\sqrt{x^2+y^2}}{\lambda z}\right)\sin\left[\frac{\pi}{\lambda z}\left(x^2+y^2\right)\right]\right\}\nonumber\\
		&\underset{k_x\lambda f,k_y\lambda f}{*}J_{1_c}\left(\frac{2\pi \rho_{\rm M}\sqrt{x^2+y^2}}{\lambda f}\right)+ \rm second\ order\ terms,
		\end{align}
and will be considered as our imaging model for the selective filtering digital in-line holography set-up.

%		\bibliography{JOSAA_Strio}
%		\bibliographystyle{ol2}

\section{Neglecting the second order term}\label{Sec:Hypothesis}
As discussed in the article main matter, our imaging model relies on a modification of the classical particle hologram formation model proposed in Ref. \cite{Tyler1976} and Eq. (\ref{Eq:Thompson}). In this model, the interference between objects are neglected assuming a diluted investigated medium. Furthermore, considering spherical objects of radius $r_i$ located at a distance $z_i$ so that
\begin{equation}\label{Eq:SimplificationThompson}
z_i\gg\frac{\pi r_i^2}{\lambda}
\end{equation}
		\begin{figure}[h]
		\centering
		\includegraphics[width = 8.4 cm]{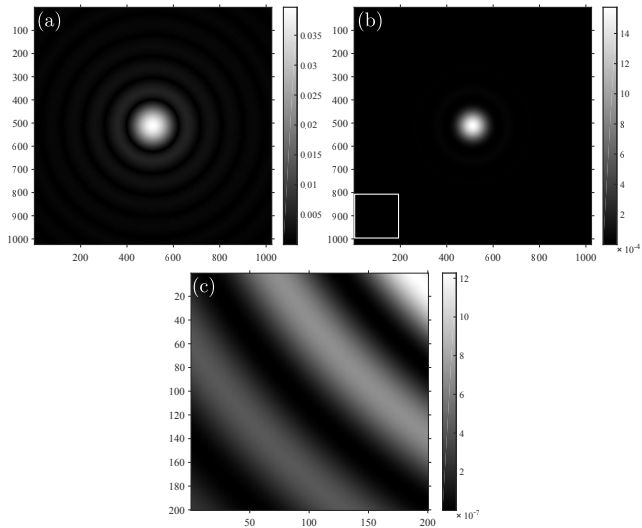}
		\caption{(Color online) (a) Representation of the first order term of Eq. (\ref{Eq:ImagingModel}). (b) Representation of the second order term of Eq. (\ref{Eq:ImagingModel}). (c) Close up view of the white square in (b).}\label{Fig:Hypothese}
		\end{figure}
the second order term of Eq. (\ref{Eq:Thompson}) can be neglected, thus leading to a simple and linear imaging model. In our proposed imaging model, the same assumption holds. It should be noted that, in our experimental configuration, the distance $z$ corresponds to the distance between the filtered image of our object through the 4f-stage to the imaging sensor.

To illustrate this aspect, both first and second order terms of Eq. (\ref{Eq:ImagingModel}) can be simulated according to our experimental parameters. As it can be noticed from Fig. \ref{Fig:Hypothese}, both the first order (a) and second order (b,c) terms span over the whole frequency space and cannot be simply canceled by our high-pass filtering strategy. However, having a closer look to the maximal values of each term, one can verify that the second order term can be neglected compared to the contribution of the reference wave and the first order term. Using Eq. (\ref{Eq:ImagingModel}) with the experimental parameters used for the experiment depicted Fig. \ref{Fig:FilteringEffect} values of each terms are found to be
\begin{align}\label{Eq:Justif}
{\rm Reference}&=\alpha^2=0.0225\nonumber\\
{\rm 1^{st} order}&=0.0253\\
{\rm 2^{nd} order}&=6.4\times 10^{-4}\nonumber.
\end{align}
As can be noticed from Eq. (\ref{Eq:Justif}), the second order term is 40 times as low as the first order term and can therefore be legitimately neglected. Moreover, both reference and first order terms are of the same order of magnitude, which confirms the result of Fig. \ref{Fig:FilteringEffect}.

	\end{document}